# Emergent quantum phenomena in two-dimensional 1T-TaS$_2$

Haiyang Chen[1], Zhiqiang Sun[1], Peng Chen[1]*

[1]Key Laboratory of Artificial Structures and Quantum Control (Ministry of Education), Tsung-Dao Lee Institute, Shanghai Center for Complex Physics, School of Physics and Astronomy, Shanghai Jiao Tong University, Shanghai 200240, China.

*Email: pchen229@sjtu.edu.cn

**Abstract**

Strong electron correlation drives 1T-TaS$_2$ from a half-filled metallic state into a Mott insulating phase, coexisting with a charge density wave at low temperatures. Under external stimuli such as pressure or ionic gating, superconductivity emerges in 1T-TaS$_2$, exhibiting an intricate relationship of competition and coexistence with the charge density wave order. In the two-dimensional (2D) limit, enhanced quantum fluctuations can stabilize a quantum spin liquid (QSL) state in the Mott insulator. This review summarizes recent advances in understanding these quantum states in 2D 1T-TaS$_2$ from the perspective of angle-resolved photoemission spectroscopy (ARPES) and scanning tunneling microscopy (STM), with a focus on the dimensionality effect on its electronic structure. We outline the signatures of QSL state in electronic spectra and discuss how this state can be revealed in the family of this material through experimental approaches beyond conventional probes such as neutron scattering. The role of Kondo effect in detecting spinon excitations is further discussed. Finally, we suggest future experimental directions and highlight how external perturbations such as gating and light excitation offer versatile pathways to control and exploit these intertwined quantum states.

# 1 Introduction

In transition metal dichalcogenides (TMDs), the diverse elemental compositions and variety of accessible chemical phases give rise to rich physics including charge density wave (CDW), superconductivity, quantum spin Hall effect, and exciton condensation [1–15]. Within this family, $TaS_2$ is a prototypical system and decades of in-depth investigations into its CDW orders and Mott insulating behavior have significantly advanced the understanding of strongly correlated electronic states. Nevertheless, the underlying driving mechanisms and emergent quantum phenomena remain elusive. With the evolution of theoretical models and improvements in experimental techniques, the microscopic picture of these collective excitations is gradually becoming clearer.

The crystal structure of $TaS_2$ is composed of S-Ta-S trilayers that are weakly bonded together by van der Waals forces, enabling the material to be thinned down to monolayer by exfoliation (Fig. 1). The intralayer hybridization between the Ta $d$ orbitals and the S $p_z$ orbital determines the formation of two primary polymorphs: the octahedral coordination structure (T-phase) and the triangular-prismatic coordination structure (H-phase), distinguished by 180° rotation of the top sulfur atomic layer relative to the bottom layer (Fig. 1a, b). Alternate stacking of H- and T-type layers can yield other polymorphs such as 4Hb and 6R, which exhibit unconventional superconductivity (Fig. 1c) [16, 17]. ARPES band structures of these phases are compared in Fig. 1d-f. 1T-$TaS_2$ is insulating with a full gap at low temperatures, whereas the 2H- and 4Hb-$TaS_2$ are metallic with bands crossing the Fermi level [18–20]. The anomalous insulating behavior of the T phase is attributed to the electron localization, commonly explained by the Mott insulator model. Notably, the H-phase is energetically more favorable than the T-phase [21]. As a result, factors such as thermal fluctuations, light irradiation, or ion implantation can drive an irreversible transition from the metastable T-phase to the H-phase [22, 23].

Single crystals of 1T-TaS$_2$ can be synthesized using the chemical vapor transport (CVT) method [24]. However, reducing the dimensionality to thin layers makes it particularly susceptible to degradation or transformation to more stable H phase, posing a significant challenge for obtaining high-quality low dimensional 1T-TaS$_2$ [25–27]. Direct growth of low-dimensional 1T-TaS$_2$ has been achieved using chemical vapor deposition (CVD) and Molecular beam epitaxy (MBE) methods (Fig. 2) [22, 28–30]. Compared to CVD, MBE operates under ultra-high vacuum and typically yields atomically flat surfaces, which are advantageous for surface-sensitive techniques such as ARPES and STM. Nevertheless, synthesizing monolayer TaS$_2$ often results in a coexistence of H and T phases [28,29]. Phase selection of TaS$_2$ can be controlled by the growth temperature and the 1T phase is favored at higher growth temperatures. Notably, increasing layer coverage will induce a partial conversion to the H-phase [22].

Bulk 1T-TaS$_2$ undergoes a first-order transition into a CDW state [31–33]. According to conventional band theory, this system is expected to be metallic. The insulating behavior observed in experiments is therefore most naturally explained by the Mott-Hubbard model [34, 35]. In this model, the kinetic energy term, governed by the hopping integral t, favors electron delocalization and band formation. However, when two electrons occupy the same site, they incur a large Coulomb energy cost U. The competition between these terms defines the system's behavior. When the interaction strength U/W (where W$\sim$nt is the bandwidth, and n is the coordination number) exceeds a critical value, it leads to the electron localization. This strong electron correlation effect splits the metallic band into upper and lower Hubbard bands, consequently opening a Mott gap. However, an ongoing debate suggests that the formation of a bilayer with a specific stacking order could also induce an insulating state [36–38], with some experiments show a clue of such a configuration [39, 40].

In Mott insulators, the dominant inter-spin interaction typically arises from superexchange, with an energy scale J$\sim$−t$^2$/U, favoring antiferromagnetic coupling. For

1T-TaS$_2$, the triangular lattice of localized spins inherently provides the necessary geometric frustration. In such a lattice, antiferromagnetic interactions between nearest neighbors cannot be simultaneously satisfied, leading to a macroscopic degeneracy of classical spin configurations. In a quantum system, this massive degeneracy is lifted by quantum fluctuations, potentially preventing conventional long-range magnetic order. Instead, spins may form a coherent superposition of entangled singlet pairs, a state famously envisioned by Anderson as a resonating valence bond (RVB) state [41]. The elementary excitations of such a QSL are spinons—charge-neutral, spin-1/2 quasiparticles that result from the spin-charge separation of electrons. These spinons can be gapped or form a gapless continuum, leading to distinct thermodynamic and spectroscopic signatures, as will be detailed below.

2D system can effectively eliminate the interlayer coupling effect from various stacking orders that complicates the situation. Moreover, the quantum fluctuations become more pronounced in lower dimensions. When it is sufficiently strong to suppress the spin ordering, a QSL state can emerge in a Mott insulator. Therefore, investigation on the CDW-Mott phase in the 2D system is crucial for understanding the rich quantum states in 1T-TaS$_2$. This review focuses on emergent quantum phenomena and related recent research in low-dimensional 1T-TaS$_2$. The article is organized as follows: Section 2 introduces the basic electronic properties of 1T-TaS$_2$, the CDW order and its chiral character; Section 3 discusses the mechanisms of the low temperature insulating phase: strong correlation driven Mott insulator or interlayer coupling induced band insulator; Section 4 provides an overview of recent advances in QSL behavior reported in monolayer 1T-TaS$_2$; finally, Section 5 concludes the review with a discussion of future experiments.

**2 Electronic properties and CDW in 1T-TaS$_2$**

Electrical transport measurements show that 1T-TaS$_2$ exhibits metallic behavior in the high temperatures and undergoes successive CDW transitions when decreasing temperature: from an incommensurate CDW (ICCDW) to a nearly commensurate CDW (NCCDW) phase at ~350 K, and further to a commensurate CDW (CCDW) phase at ~180 K, as shown in Fig. 3a [10, 31]. In the low temperature CCDW phase, the system becomes insulating. Within this phase, thirteen in-plane Ta atoms distorted into a Star of David (SOD) to form a ($\sqrt{13} \times \sqrt{13}$) superstructure with one unpaired electron localized on the center Ta atom [33, 42], where the Mott state is simultaneously developed. The superlattice is rotated by 13.9° relative to the underlying atomic lattice [32]. In the NCCDW phase, commensurate domains form into hexagonal clusters separated by discommensurate areas.

The CDW superstructure is arranged in a triangular lattice, which breaks the in-plane mirror symmetry, giving rise to two energetically degenerate chiral configurations with opposite handedness (Fig. 4a) [18]. These two chiral structures can be distinguished using chirality dependent Raman response to left-handed (σ+) and right-handed (σ-) circularly polarized light (Fig. 4c). The key signature is the inequality of Raman intensities in cross-circular polarization configurations ($I_{\sigma+\sigma-} \neq I_{\sigma-\sigma+}$) [18, 43]. Specifically, the in-plane Eg phonon modes exhibit distinct intensities under opposite polarizations, confirming the in-plane chiral nature of the CDW structure [18]. Additional evidence for chirality comes from second-harmonic generation (SHG), a sensitive probe of inversion symmetry breaking. 1T-$TaS_2$ exhibits a clear difference in SHG intensity when excited with the σ+ versus σ− circularly polarized light (Fig. 4b). Furthermore, chiral electronic band structure, characterized by windmill-shaped Fermi surfaces is revealed by ARPES [18]. The broken time-reversal symmetry and orbital order in chiral CDW may stabilize a chiral superconducting state, a scenario theoretically predicted in systems such as Kagome superconductors [44].

The observed planar CDW chirality in $TaS_2$ is shown to be linked to a ferro-rotational order defined by a structural axial-vector order parameter, where the two chiral domains correspond to globally clockwise and counterclockwise rotations of the SOD lattice [45]. Recent experiments have achieved non-volatile electrical switching between these ferro-rotational domains in nanometer-thick samples [46]. This switching is remarkable because the ferro-rotational order parameter is invariant under both time reversal and spatial inversion, preventing direct coupling with an external electric field. It is achieved indirectly by the field-driven motion of domain walls [46]. This hysteretic and electrically controllable bistability of the chiral/rotational state demonstrates its potential for device applications such as non-volatile memory.

The CDW state in 1T-TaS$_2$ is highly sensitive to external perturbations. It is shown the phase transition can be driven reversibly by electric fields or laser excitation, attributed to a carrier-induced collapse of the gap [47, 48]. Optical pumping can also induce a hidden or metastable CDW state (H-CDW), and rapid thermal quenching of the sample enables reversible switching between the H-CDW and C-CDW phases [49–52]. STM studies of quenched samples reveal the formation of nanoscale domain structures [53]. Despite the metallic nature of the domain walls, the overall electrical behavior of the sample remains insulating—a phenomenon explained by carrier localization effects due to CDW stacking disorder [54].

The CDW transition persists down to the 2D limit, as demonstrated by the observation of $\sqrt{13} \times \sqrt{13}$ superstructure in STM topographic images of monolayer 1T-TaS$_2$ (Fig. 5a) [22]. ARPES spectra of monolayer TaS$_2$ film show a flat band (labeled as α) centered around the $\bar{\Gamma}$ point (Fig. 5b) and the 1T system is insulating with a gap of 0.26 eV. However, such a gap cannot be attributed to the formation of CDW because conventional band theory fails to account for it [30, 55], pointing to the essential role of electron correlation effects. The CDW-related gaps are observed in the deeper binding energies of 0.3 and 0.9 eV and separate the γ band from Ta 5$d$ state into several pieces [30]. These gaps are closing with increasing temperature. CCDW phase in monolayer and few-layer 1T-TaS$_2$ grown via CVD or MBE are more robust, with a transition temperature (~230-353 K) significantly higher than that of bulk [30, 56, 57]. The dimensionality-dependent CDW correlation length has been determined to be approximately 7.8 nm [56, 58].

The intricate relationship between CDW and emergent superconductivity constitutes a central theme in 1T-TaS$_2$. While the pristine bulk material is insulating at low temperatures, superconductivity can be induced by perturbations that suppress the Mott-CCDW state and enhance metallic character. Key pathways include applying hydrostatic pressure to broaden the bandwidth, chemical intercalation or ionic gating to inject carriers, and exploiting chemical substitutions [10,57,59,60]. These methods typically lead to a superconducting dome appearing adjacent to the suppressed CCDW phase, suggesting a competitive relationship. However, evidence of NCCDW phase coexists with superconductivity indicates a more complex, potentially intertwined interplay. It has been proposed that superconductivity may nucleate specifically within the metallic interdomain spaces of the NCCDW phase [10], suggesting a real-space microscopic phase separation between insulating CCDW domains and superconducting regions. This scenario is further supported by the observed analogy to cuprate superconductors where electronic phase separation (e.g., stripes) is crucial [61]. However, this model contrasts with scenarios where superconductivity arises from a self-doped Mott phase or is mediated by CDW fluctuations [62,63], as the superconductivity in 1T-TaS$_2$ shows remarkable insensitivity to the collapse of the CDW state. Thus, 1T-TaS$_2$ serves as a pivotal platform for exploring the potential of nanoscale electronic phase separation as a pathway to stabilize superconductivity proximate to insulating orders.

## 3 A Mott insulator or a band insulator?

The low-temperature insulating behavior of 1T-TaS$_2$ is unexpected because of the formation of SOD in the CDW state and presence of one unpaired electron localized on the center Ta atom in $\sqrt{13} \times \sqrt{13}$ superstructure, which gives rise to a metallic band crosses the Fermi level. This picture is supported by neutron scattering that a magnetic moment of ~0.4 μB per SD cluster is estimated [64]. Mott insulator is the widely accepted explanation that strong electron correlations localize electrons and open a Coulomb gap. This effect splits the metallic band into upper and lower Hubbard bands, as illustrated in Fig. 6a. Experimental evidences include two separated peaks around the Fermi level and an insulating gap of ~420 meV observed in STS spectra [65]. ARPES measurements on bulk 1T-TaS$_2$ show a "flat band" near the Fermi level, attributed to the lower Hubbard band [66–68]. This band is dispersive in energy along the out-of-plane momentum, demonstrating the interlayer interaction tends to enhance the inter-site hopping and broaden the bandwidth of the Hubbard band [69]. The upper Hubbard band has been revealed using pump-probe techniques and inverse-photoemission spectroscopy [70, 71].

An alternative explanation for the insulating state involves interlayer coupling effects under different stacking configurations [36–40]. As shown in Fig. 7, the relative orientation of the orbitals in adjacent *ab*-planes has a significant effect on band dispersions [36]. Density functional theory calculations predict a gap around the Γ point but the band structure remains metallic along out-of-plane momentum direction when the SOD central atoms in adjacent layers are aligned along the *c*-axis (A-stacking). Further calculations reveal a full gap when a paired stacking shifts by half a unit cell every two layers (AL-stacking) [37, 69]. The formation of a gap is due to the interlayer hybridization mostly mediated by the $d_{3z^2-r^2}$ orbitals. These calculations were performed without an on-site repulsion U on Ta, arguing against Mott physics as the origin of the gap. Synchrotron-based X-ray diffraction (XRD) shows the half-integer-indexed diffraction peaks along the *c* axis in bulk 1T-TaS$_2$, providing evidence of dimerization along this direction [68].

STM studies on the bulk material surfaces have observed the dominant "large-gap" spectra (~0.4–0.45 eV), "small-gap" spectra (gap ~0.1–0.2 eV) and occasionally metallic spectra in certain regions [39, 72, 73]. Stacking order was determined from the single-step areas between neighboring layers. It turns out different stacking orders exhibit similar large-gap spectra when the measurements are performed away from the step edge, suggesting the large insulating gap in the bulk is independent of stacking order and supports a correlation-induced Mott-Hubbard mechanism. Note that the misaligned stacking near step edges may disturb the Mott state to a small-gap or metallic state, likely due to the defects or disorder effects [72]. Electron doping has been used to study the different behaviors expected for band and Mott insulators [40], based on their distinct responses due to the fundamental difference of the half- and full-filled orbitals involved. Alkali metal deposition on 1T-TaS$_2$ surfaces reveals that STS measurements on the paired bilayer surface shows a rigid shift expected for the doping of an ordinary band insulator. In contrast, STS on the surfaces with half-filled Ta $5d_{3z^2-r^2}$ states shows suppression of the upper Hubbard band.

Similar stacking order effects have been revealed in isostructural compound 1T-TaSe$_2$ [74]. A key difference between 1T-TaSe$_2$ and 1T-TaS$_2$ is that stacking order in 1T-TaSe$_2$ induces more substantial perturbation. While 1T-TaSe$_2$ exhibit a large spectral gap on the surface, it shows metallic behavior in transport experiments at low temperatures [74–76]. This discrepancy may arise from connected metallic areas of electrodes in the transport measurements. Recent ARPES measurements with micro-size light spot directly show the metallic and insulating band structures coexist in different domains on the surface caused by the different stacking arrangement [77].

Although dimerization of 1T-TaS$_2$ with specific stacking indicates a similar electron structure to a band insulator, it cannot explain why monolayer system—interlayer coupling and stacking order effects are absent—show a large-gap insulating state. The Mott physics thus provides the most natural explanation. STS spectra taken from MBE-grown monolayer 1T-TaS$_2$ reveal features similar to the large-gap spectra of bulk materials [23]. Differential conductance spectra show a clear gap of 0.45 eV near the Fermi level, surrounded by sharp Hubbard peaks (Fig. 6d). These results agree well with the first-principles calculations that the Hubbard peaks are dominated by the density of states of the central Ta atoms (Fig. 6b). In these calculations, a Hubbard U is included to account for the Coulomb interaction of Ta $5d$ orbitals. The existence of localized spins in SODs, consistent with the Mott picture, has been confirmed by the observation of Kondo resonance peaks around the Fermi level in monolayer 1T-TaS$_2$ on top of metallic 1H-TaS$_2$ [78].

As shown in Fig. 8a, ARPES spectra of monolayer 1T-TaS$_2$ display a flat band (lower Hubbard band) near the Fermi level and a spectroscopic gap of 0.26 eV, determined from the energy difference between the lower Hubbard band and the Fermi level [30, 79]. Temperature-dependent ARPES measurements track the gradual closing of the Mott gap with increasing temperature, allowing extraction of a CDW-Mott transition temperature $T_C$ = 353 ± 12 K (Figs. 8b and 8c) [30]. It is approximately twice the value observed in bulk material (~180 K), demonstrating the robustness of the Mott phase in 2D 1T-TaS$_2$. The temperature-dependent gap shows no obvious difference between the cooling and heating processes in monolayer samples indicates the hysteresis observed in the bulk [31, 32] likely originates from competing stacking orders in the 3D system.

As a summary, while interlayer stacking effects can induce a gap in specific configurations, the correlation-driven Mott scenario provides a more comprehensive and natural explanation for the insulating behavior in 1T-TaS$_2$, particularly in its monolayer limit. The key experimental signatures expected for the two pictures exhibit systematic differences: the Mott gap arises from on-site Coulomb repulsion, manifesting in spectroscopic data as separated Hubbard bands (a lower and an upper band) around the Fermi level. However, the insulating gap in the stacking picture originates from interlayer hybridization in specific stacking geometries, which modifies the band dispersion without necessarily generating distinct Hubbard band features. A Mott insulator involves a half-filled orbital, leading to a non-rigid shift and significant spectral weight redistribution of its Hubbard bands upon doping (e.g., suppression of the upper Hubbard band) [40]. In contrast, a stacking-induced band insulator involves filled/empty bands, and its doping response resembles a more rigid shift of the entire band structure.

## 4 Quantum spin liquid behavior in monolayer 1T-TaS$_2$

QSLs represent a novel class of quantum matter characterized by highly entangled spins that fail to develop a long-range magnetic order even at 0 K [80]. One of the most exotic phenomena predicted in QSLs is the emergence of fractionalized excitations leading to the separation of the spin and charge components of an electron. Despite intense study since the concept was introduced by P.W. Anderson in 1973 [41], the experimental identification of QSLs remains elusive, as no "smoking gun" evidence can provide definitive conclusion. Mott insulators offer a promising platform for realizing a QSL state when long-range spin order is suppressed, which can be realized through frustration and/or quantum fluctuations. Frustration creates a massively degenerate manifold of competing ground states, while quantum fluctuations act to destabilize any incipient order.

1T-$TaS_2$ has been proposed as a gapless QSL candidate [81]. The SOD clusters formed at low temperatures are arranged in a triangular lattice, which exhibits strong geometric frustration. While the Mott mechanism behind its insulating ground state is still debated, subsequent experiments on 1T-$TaS_2$ suggested to be consistent with QSL scenario. For instance, magnetic susceptibility of bulk 1T-$TaS_2$ shows a nearly temperature-independent Pauli paramagnetic susceptibility, and muon spin relaxation measurements reveal the absence of any signs of static magnetic order or spin freezing down to 20 mK, as shown in Figs. 9a and 9b [64, 82]. Low-temperature specific heat measurements reveal a linear term $\gamma T$ and thermal conductivity exhibits the finite residual linear term in the zero-temperature limit (Fig. 9c), consistent with the presence of gapless spinons with a Fermi surface [83, 84]. Nuclear quadrupole resonance provides a local probe of electronic spin fluctuations via nuclear spin-lattice relaxation measurements in zero field, showing gapless QSL dynamics in 1T-$TaS_2$ (Fig. 9d) [85]. However, some controversial results exist, including report of no discernible residual linear term of thermal conductivity at zero field [86]. These discrepancies are attributed to differences in sample quality, such as varying degrees of disorder, defects, or stacking faults.

High quality monolayer $TaS_2$, free from the interlayer stacking issue, is ideal for studying the potential QSL behavior. Furthermore, enhanced quantum fluctuation as reduced dimensionality can help stabilize such a state. However, experimental verification remains a major challenge because traditional bulk-sensitive techniques such as neutron scattering, specific heat measurements, and nuclear magnetic resonance have difficulties when applied to ultra-thin samples due to insufficient signal strength, and dominant substrate interference. The development of surface sensitive and local-probe approaches is therefore essential for uncovering signatures of QSLs in 2D materials.

In a QSL, electrons fractionalize into spinons and chargons with independent of the spin and charge components, respectively. When an electron is removed from a QSL, the resulting hole state separates into a spinon hole and a holon. Within a mean-field approximation, the convolution of their dispersions can be expressed as the electronic spectral function, resulting in a characteristic composite spectrum that should be readily probed by ARPES (Figs. 10a and 10c) [87]. A similar process occurs in STM measurement, where the tunneling rate of an electron at a particular energy can be expressed as a convolution of spinons and chargons. An STM study on monolayer 1T-TaSe$_2$ revealed long-wavelength super-modulations at the energies of lower and upper Hubbard bands [88], consistent with the predictions of spinon density modulation arising from a spinon Fermi surface instability [81, 89]. Similar STM results have been reported for monolayer 1T-NbSe$_2$ [90].

Recently, an ARPES study provided spectroscopic evidence supporting a QSL state in monolayer 1T-TaS$_2$ grown by MBE method [30]. At low temperatures, the observations of a flat band below the Fermi level and a gap opening of about 200 meV demonstrate 2D Mott insulating nature in the system. This flat band exhibits several "unusual" features: it is nearly dispersionless with an extremely narrow width (~10 meV), yet shows anomalous broadening in momentum space, and the spectral weight decays rapidly with increasing momentum. In contrast, conventional band theory predicts a single and continuous band around the Fermi level. This observation is interpreted as a spectral continuum arising from electron fractionalization. Spectral function simulation results based on a low-energy effective model agree well with the ARPES spectra (Figs. 10d and 10e).

Directly probing the spinon Fermi surface is challenging due to its charge-neutral character. As illustrated in Fig. 11, the Kondo effect involving spinons can give rise to new peaks near the edges of Hubbard bands [90, 91]. Such resonance peaks have been observed via STS in monolayer 1T-TaSe$_2$ and 1T-NbSe$_2$ by depositing magnetic atoms on the surfaces [91]. When itinerant spinons couple to the magnetic impurities, the formation of a spinon Kondo cloud can induce detectable resonance peaks through emergent gauge-field fluctuations. Furthermore, the coupling between spin impurities-spinons and the chargons will lead to a charge redistribution and influence the Mott gap. With doping magnetic impurities (Fe or Co) on the monolayer 1T-TaS$_2$ surface, ARPES measurements show the reduced intensity of the flat band and some spectral intensity filling up the Mott gap around the lower Hubbard band edge. In contrast, non-magnetic impurity (Na or K) doping primarily induces a rigid chemical potential shift [30]. Similar doping effects have been observed by ARPES in other monolayer 1T-MX$_2$ (M = Ta, Nb; X = Se, S) [55].

Among these experimental observations discussed, the absence of static magnetic order down to 20 mK (as established by µSR) serves as a robust prerequisite, which confirms that the system remains in a dynamic, highly entangled state far below its interaction energy scale, ruling out conventional magnetic ground states. Highly suggestive, yet more model-dependent, signatures come from bulk thermodynamic and transport measurements. The most intriguing and direct spectroscopic evidence for fractionalization emerges from monolayer studies, particularly the ARPES observation of an anomalous spectral continuum and STM signatures of spinon density modulations, representing a newer frontier whose interpretation within specific QSL frameworks would benefit from further theoretical consolidation.

## 5 Conclusion and outlook

Recent studies on bulk and monolayer 1T-TaS$_2$ have revealed a rich landscape of quantum phenomena, arising from the complex interplay of CDW order and Mottness in the ground state. The proposed QSL state in the Mott phase of monolayer 1T-TaS$_2$ is supported by converging experimental evidences, including spectroscopic signatures of electron fractionalization observed via ARPES, local electronic structure studies through STM/STS, and distinctive responses to magnetic impurity doping. These studies open up new avenues for investigating QSL physics from the perspective of electronic degrees of freedom.

While the study of monolayer 1T-TaS$_2$ eliminates the variable of interlayer stacking order, it introduces other extrinsic factors whose potential influence warrants careful consideration. Disorder—arising from point defects, sulfur vacancies, or domain boundaries inherent to the growth and transfer processes—can locally perturb the electronic structure. In a Mott insulator, disorder can create in-gap states, leading to a spurious "filling" of the spectroscopic gap in STS measurements. Strain can collapse the Mott gap and drive an insulator-metal transition in 1T-TaS$_2$ [92]. Another significant factor is the substrate coupling. Even van der Waals substrates can induce charge transfer, modify dielectric screening, and impose symmetry-breaking potentials. As an example, increased screening strength in HOPG substrate slightly reduces the gap in monolayer 1T-TaSe$_2$ [88]. However, an additional 2 × 2 super-modulation (with respect to the CDW lattice) has been revealed with HOPG substrate, which is attributed to the higher harmonics of the Spinon Fermi surface instability wavevectors [93].

The rich phase diagram and tunability of 1T-TaS$_2$ present a vast landscape for future exploration. Building on the recent demonstration of ferro-rotational domain switching by electric field [46], future work can explore the role of disorder in switching fidelity, and gating experiments in monolayer devices to continuously tune the Mott gap and probe the quantum critical point. The electrically-driven switching between insulating and metallic states can be exploited for ultra-low-power non-volatile memory. Time-resolved studies can disentangle the intertwined degrees of freedom based on distinct characteristic timescales for electron-phonon interactions and electron-electron correlations [94].

Therefore, ultrafast pump-probe method is intriguing to reveal the relationship between Mott phase and CDW orders in 1T-TaS$_2$ and track the non-equilibrium dynamics.

Another promising direction has emerged with the development of van der Waals heterostructures based on 1T-TaS$_2$. The artificial stacking of 1H/1T-TaS$_2$ bilayer heterostructure has successfully demonstrated the realization of a heavy-fermion system in two dimensions [78]. This is achieved by engineering the Kondo coupling between the lattice of localized magnetic moments in the Mott-insulating 1T layer and the itinerant electron sea in the metallic 1H layer. Given that strongly correlated systems such as Mott insulators and heavy-fermion compounds are fertile ground for unconventional superconductivity, the tunable nature of 1T-TaS$_2$-based heterostructures provides a promising platform for engineering such superconducting states through controlled dimensional tuning and proximity effects.

Nano- or micro-focused ARPES is ideally suited for detecting the local CDW domains and link the local structure to the electronic structure. By focusing the photon beam to a sub-micron spot, this technique can selectively probe electronic bands from individual domains or at domain walls. This would provide a direct, momentum-resolved answer to whether the proposed metallic interdomain regions in the NCCDW phase host a distinct electronic signature that differs from the insulating domains. Furthermore, it could map the electronic structure across intentionally engineered heterointerfaces (e.g., 1T/1H stacking regions), revealing the momentum-space footprint of proximity effects and interlayer coupling.

**Acknowledgements**

We thank Prof. Xiaoqun Wang and Prof. Xiaoyan Xu for helpful discussions.

**Author contributions**

H. Y. C. and Z. Q. S. drafted the first version of manuscript. P. C. conceived the work

and revised the manuscript. All authors read and approved the final manuscript.

**Funding information**

The work at Shanghai Jiao Tong University is supported by the Ministry of Science and Technology of China under Grant No. 2022YFA1402400, the National Natural Science Foundation of China (Grant No. 12374188).

**Availability of data**

The supporting information of this study are available from the corresponding author upon reasonable request.

**Declarations**

**Competing interests**: The authors declare no competing interests.

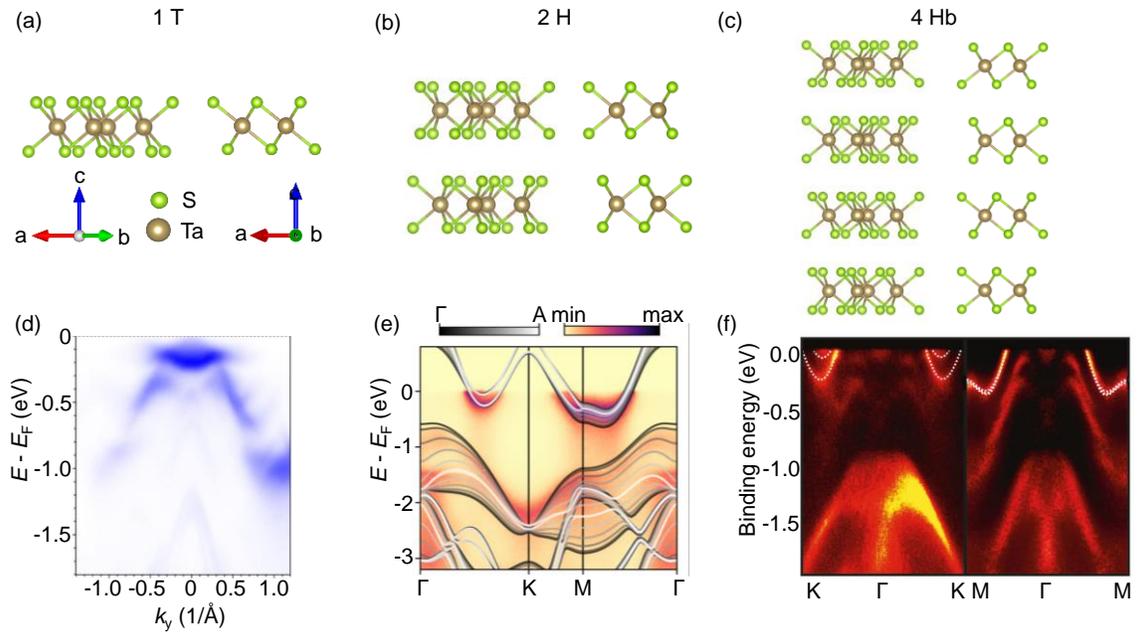

**Figure 1** (a-c) Atomic structures of 1T-, 2H- and 4Hb-TaS$_2$. (d-f) A comparison of ARPES spectra of bulk T-phase, H-phase and 4Hb-TaS$_2$. (Figures adapted from refs. [18–20])

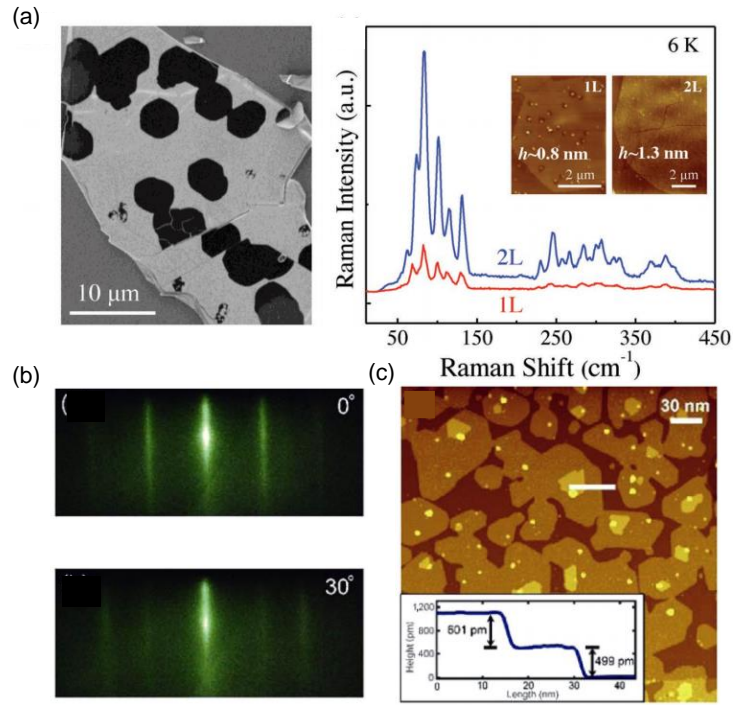

**Figure 2** (a) SEM image of 1T-TaS$_2$ grown on *h*-BN. (b) Raman spectra taken from monolayer and bilayer 1T-TaS$_2$ samples. Insets are corresponding AFM images. (c) RHEED patterns of the monolayer TaS$_2$ thin film along two incident azimuth angles of 0° and 30°. (d) An STM topographic image shows a coexistence of monolayer and bilayer TaS$_2$. (Figures adapted from refs. [22, 24])

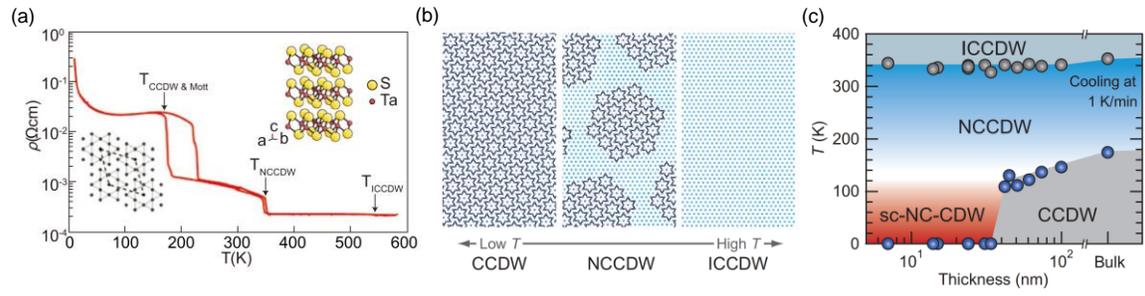

**Figure 3** (a) Phase transitions of 1T-TaS$_2$ with temperature measured by transport. (b) Schematic pictures of a Ta atom network in the CCDW (left), hexagonal NCCDW (middle), and ICCDW (right) phases. (c) Temperature-thickness phase diagram of 1T-TaS$_2$ nano-thick crystals. (Figures adapted from refs. [10, 42])

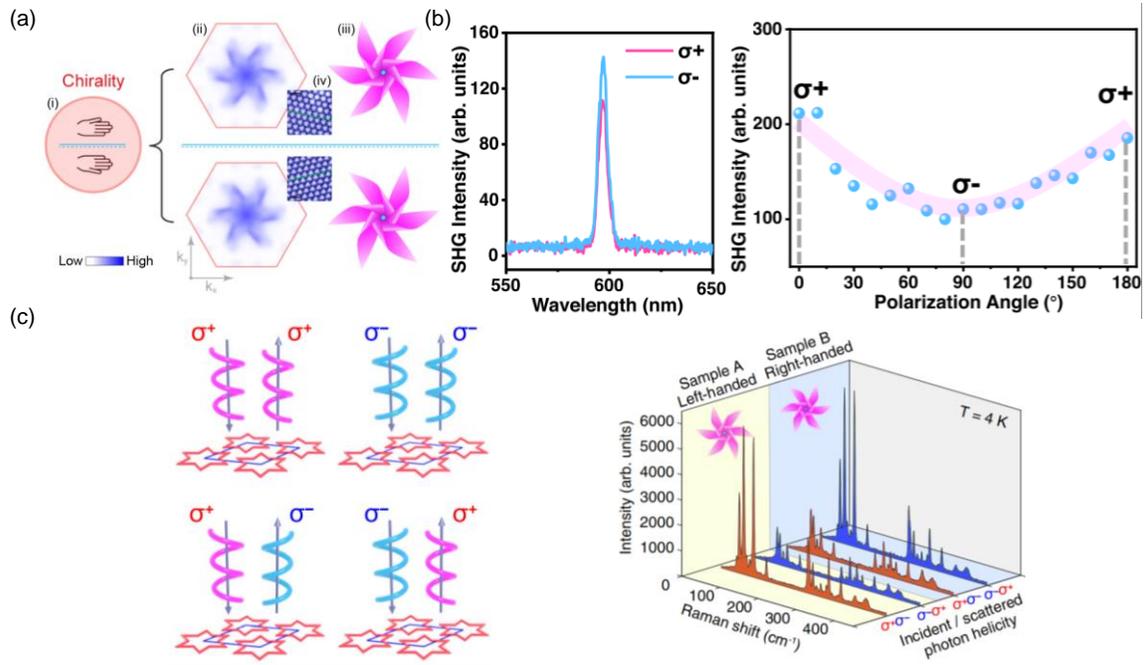

**Figure 4** (a) The hidden chiral order in 1T-TaS$_2$, which is featured by the chiral CDW and windmill-shape band structure with broken mirror symmetry. (b) SHG signals of 1T-TaS$_2$ samples and variation of SHG intensity as a function of the polarization angle. (c) Schematic of polarization-resolved Raman response measurements and Raman spectra of samples with left- or right-handed domains showing different $\sigma^+\sigma^-/\sigma^-\sigma^+$ response. (Figures adapted from refs. [18, 25, 43])

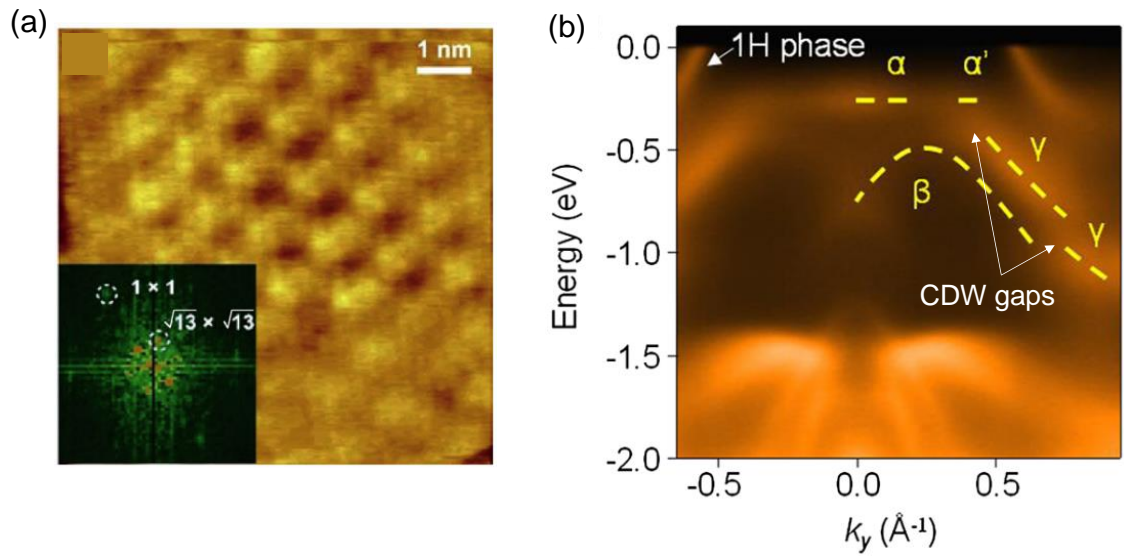

**Figure 5** (a) Atomically resolved STM image of the 1T-TaS$_2$. The inset is the fast Fourier transform result. (b) ARPES maps taken along $\overline{\Gamma M}$ with 45 eV $p$ polarized light at 10 K. CDW gaps are marked with arrows. (Figures adapted from refs. [22, 30])

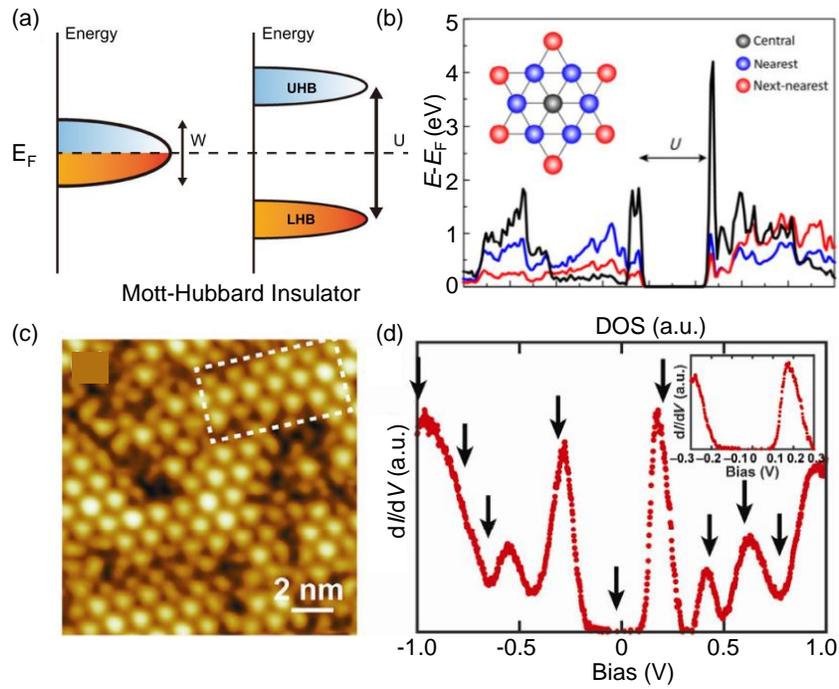

**Figure 6** (a) Schematic diagram of the Mott-Hubbard Insulator. (b) PDOS of the three types of nonequivalent Ta atoms in an SOD for monolayer 1T-TaS$_2$. (c) The topographic image of monolayer TaS$_2$ with randomly distributed defects. (d) The large range d$I$/d$V$ spectrum obtained on the center of an SOD. (Figures adapted from refs. [10, 23])

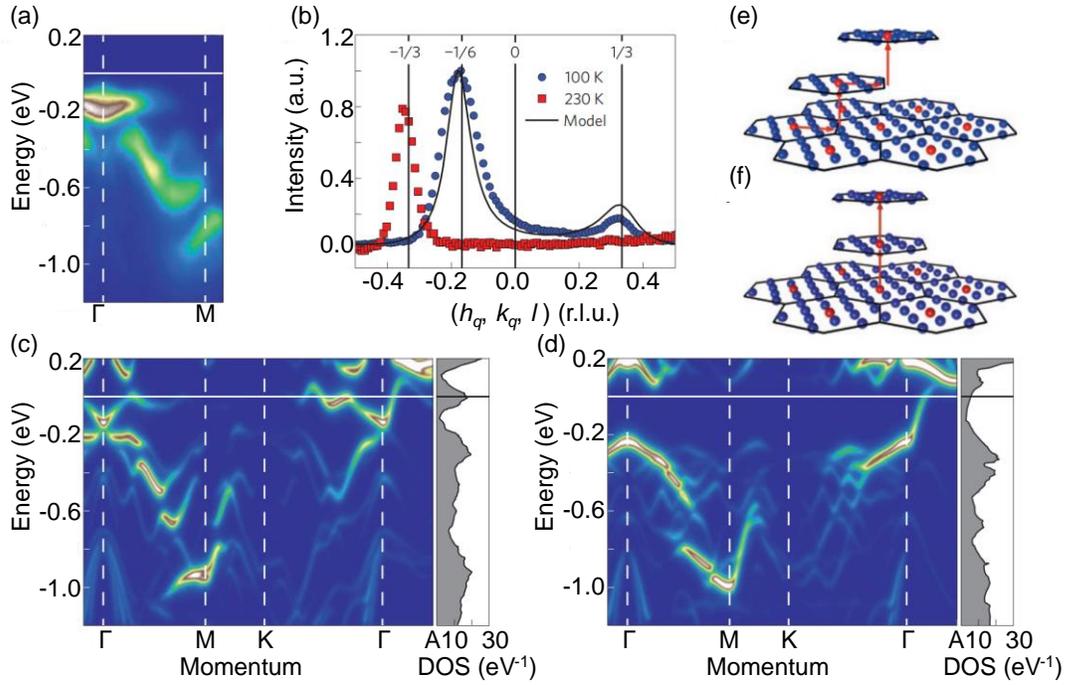

**Figure 7** (a) ARPES data of 1T-TaS$_2$ measured at 1 K. (b) XRD intensity of 1T-TaS$_2$ measured along the *l*-direction. (c-d) Calculated band structure for T$_S$ = 2**a** + **c** and T$_S$ = 1**c**, respectively. **a**, **c** are the lattice vectors of the undistorted structure. (e-f) Visualization of the stacking with T$_S$ = 2**a** + **c** and T$_S$ = 1**c**, respectively. Red arrows indicate the stacking vectors that connect the central Ta-sites (shown in red) in successive layers. (Figures adapted from ref. [36])

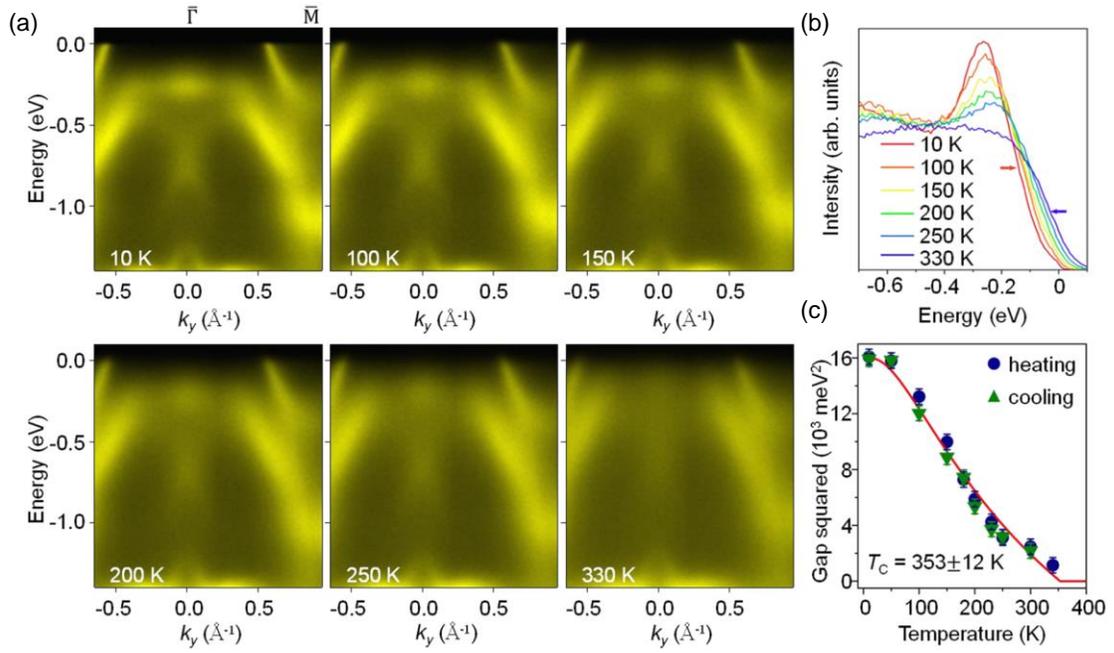

**Figure 8** (a) Temperature dependence of ARPES spectra of the band structure and the energy gaps. (b) EDCs at the zone center at selected temperatures. (c) The extracted temperature dependence of the square of the energy gap. (Figures adapted from ref. [30])

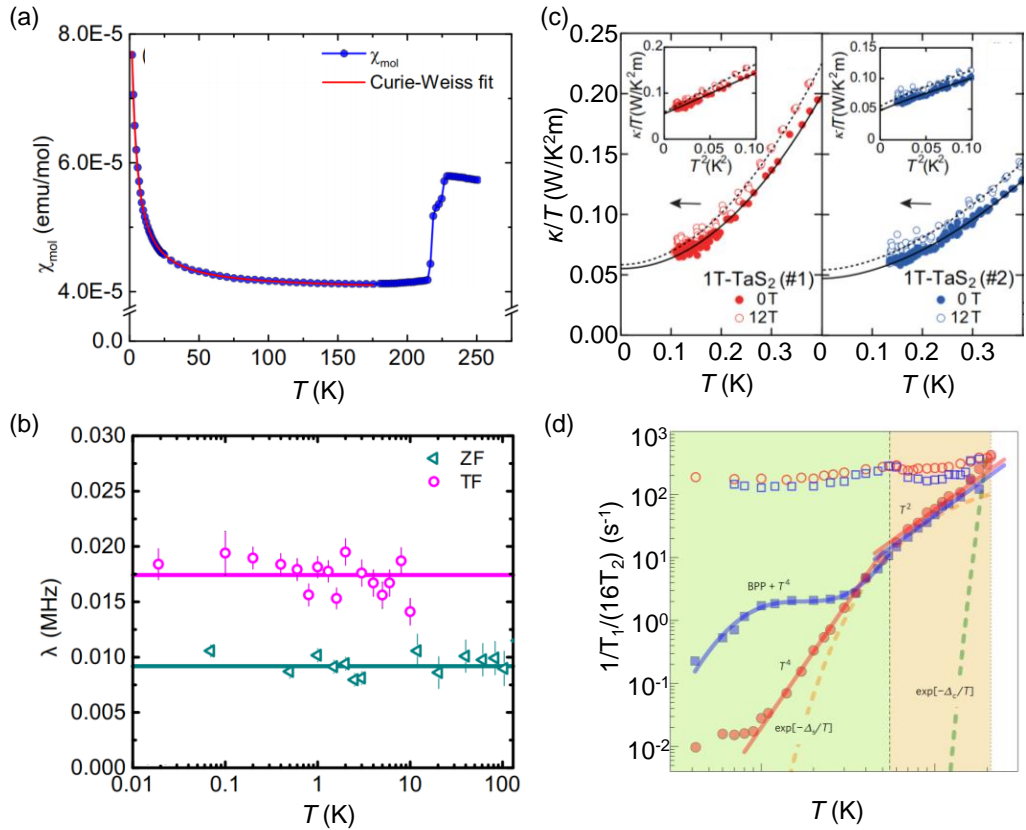

**Figure 9** (a) Magnetic molar susceptibility vs temperature, obtained from SQUID magnetization measurements at constant magnetic fields of 1 and 3 T. (b) Temperature dependence of the muon spin damping rate $\lambda$. (c) Temperature dependences of $\kappa/T$ for 1T-TaS$_2$ in zero magnetic field and at $\mu_0 H$ = 12 T applied along $c$ axis. (d) The temperature dependences of the Ta spin-lattice relaxation rate $1/T_1$ and of the Ta spin-spin relaxation rate $1/T_2$ for the Ta $\alpha$ and $\beta$ sites. (Figures adapted from refs. [83–85])

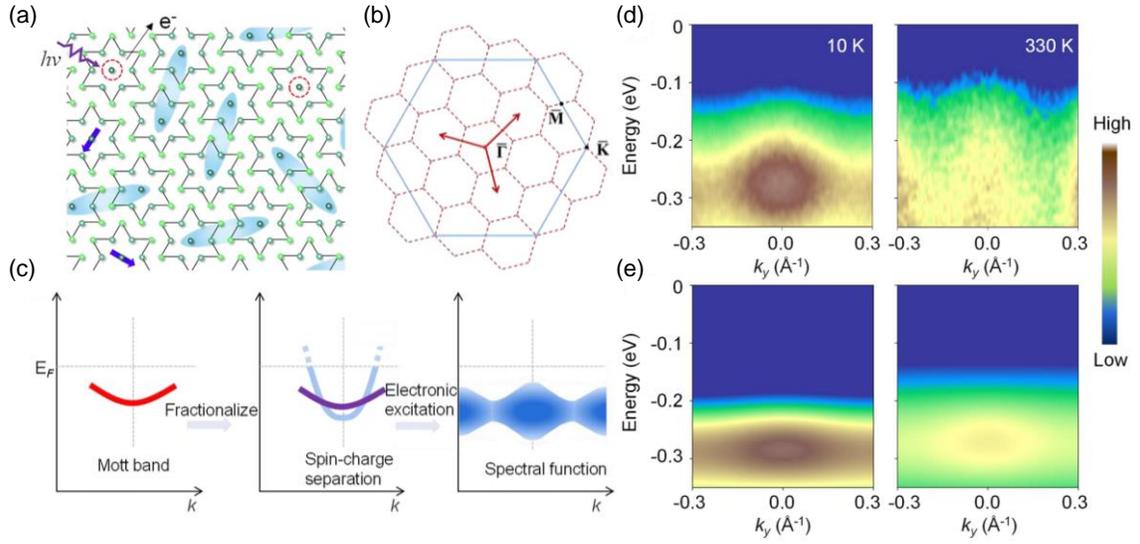

**Figure 10** (a) A schematic diagram of excitations in the QSL state. (b) Brillouin zones of the (1 × 1) and ($\sqrt{13} \times \sqrt{13}$) structures are outlined in blue and red, respectively. (c) Schematic of the electron fractionalization process and spectral dispersion after convolution in monolayer 1T-TaS$_2$. (d) ARPES spectra taken along the $\overline{\Gamma M}$ direction at 10 and 330 K. (e) Spectral function simulation results based on a low energy effective model at 10 and 330 K. (Figures adapted from ref. [30])

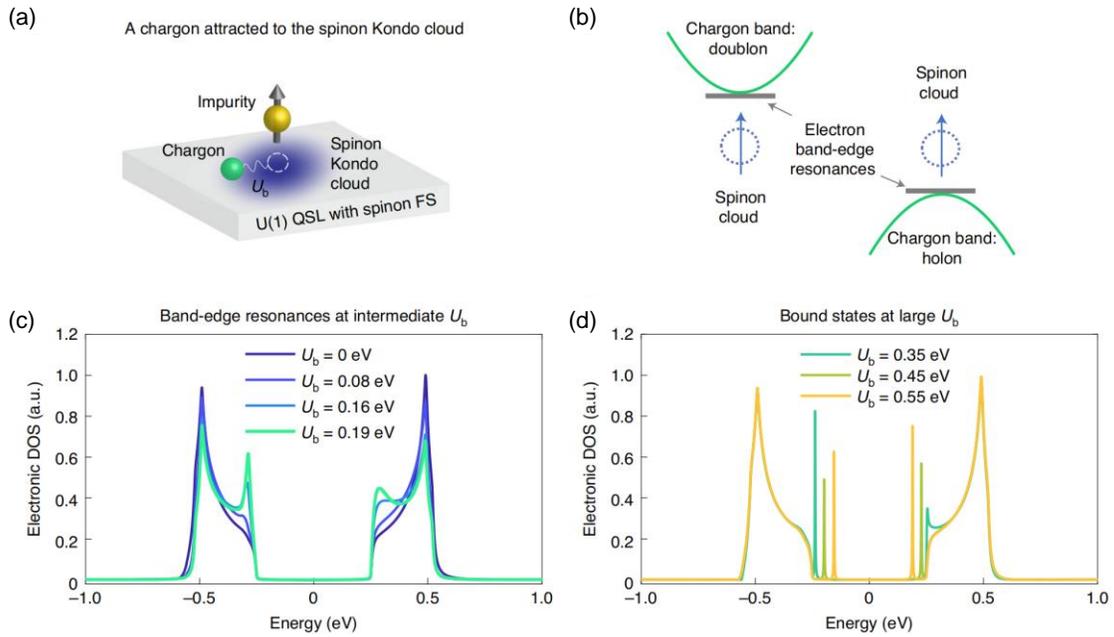

**Figure 11** (a) Sketch of chargon binds to the spinon Kondo cloud. (b) Schematic diagram of formation of spinon-charron band-edge resonance states. (c) With increasing binding interaction $U_b$, resonance peaks emerge at the Hubbard band edges. (d) With larger $U_b$, the spinon–chargon resonance states move inside the Mott gap and exhibit narrow peaks. (Figures adapted from ref. [91])